\documentclass[%
 reprint,
 onecolumn,
 longbibliography,
 amsmath,amssymb,
 aps,
]{revtex4-2}

\usepackage{lipsum}
\usepackage{color}
\usepackage{natbib}
\usepackage{graphicx}% Include figure files
\usepackage{dcolumn}% Align table columns on decimal point
\usepackage{bm}% bold math
\begin{document}

%\preprint{APS/123-QED}

\title{Prandtl-Tietjens intermittency in transitional pipe flows}

\author{Rory T. Cerbus}
\email{rory.cerbus@u-bordeaux.fr}
\affiliation{
Université de Bordeaux \& CNRS, LOMA (UMR 5798), 33405 Talence, France}
% \homepage{https://www.loma.cnrs.fr/rory-cerbus/}


\date{\today}% It is always \today, today,

\begin{abstract}
%\lipsum[1-1]
Pipe flow often traverses a regime where laminar and turbulent flow co-exist. Prandtl and Tietjens explained this intermittency as a feedback between the fluctuations of the internal flow resistance and the constant pressure drop driving the flow. However, because the focus has moved towards studying intermittency without flow fluctuations near the universal critical Reynolds number, their explanation has largely disappeared. Here we refine the mechanism, which has never been put to a quantitative test, to develop a model that agrees with experiments at higher Reynolds numbers, enabling us to demonstrate that Prandtl and Tietjens’ mechanism is, in fact, intrinsic to flows where both the pressure gradient and perturbation are constant.
\end{abstract}

\maketitle

%\tableofcontents

%\section{\label{sec:level1}First-level heading:\protect\\ The line
%break was forced \lowercase{via} \textbackslash\textbackslash}

%\subsection{\label{sec:level2}Second-level heading: Formatting}

In 1839, while investigating the friction in pipe flow, Gotthilf Hagen \cite{hagen1839ueber} observed that the jet of water exiting the pipe resembled a glassy rod at low flow speeds, which then began to pulse as the flow speed increased. The jet reflects the state of the flow inside the pipe. It is in one place glassy \cite{hagen1839ueber} and smooth \cite{couette1890distinction}, ``laminar", while frosty \cite{couette1890distinction} and sinuous \cite{reynolds1883an}, ``turbulent", elsewhere. Hagen's pulses were a manifestation of this intermingling of laminar and turbulent flow, which we now call \emph{intermittency}, a basic feature of the transition to turbulence in pipe flow and other shear flows \cite{mullin2011experimental,eckhardt2007turbulence,letellier2017intermittency,tasaka2010folded,avila2011onset}. The turbulent patches, which can also die, split, or grow, are carried downstream so that the whole pattern of intermittency changes continuously in space and in time. 
The phenomenon of intermittency was unexpected, given that the flow conditions were kept as constant as practical, and its origin was at first unclear \cite{hagen1839ueber,couette1890distinction,sackmann1954changements,reynolds1883an}. In their famous fluid mechanics textbook, Prandtl and Tietjens \cite{tietjens1957applied}, hereafter referred to as PT, qualitatively explained intermittency as the result of a feedback between the larger friction in the turbulent patches and the constant total pressure drop driving the flow \cite{tritton2012physical}. With a larger friction, the flow speed decreases until it is reduced below the critical speed, so that no new turbulence is created. When the increased friction of the patch leaves the pipe, the flow speed increases. The critical speed is again exceeded, a new patch is created, and the cycle repeats. In the PT mechanism, intermittency not only creates but requires fluctuations in flow speed, both of which oscillate. In keeping with common practice, we will hereafter use the non-dimensional flow speed or Reynolds number, $Re = UD/\nu$, where $U$ is the flow speed, $D$ is the diameter, and $\nu$ is the kinematic viscosity.

The qualitative PT mechanism remained the prevalent explanation until the seminal study of transitional pipe flow by J. Rotta \cite{rotta1956experimenteller}. Rotta accepted the general validity of the PT mechanism but sought to determine if it was the only source of intermittency by taking great pains to maintain an approximately constant $Re$ in his constant pressure drop and constant perturbation experiments \cite{rotta1956experimenteller}. He introduced a large external resistance into his pipe system so that the pressure drop over this resistance would damp out oscillations. Restricting attention to $Re \lesssim 3000$, he found that the intermittency persisted, despite no obvious fluctuations in $Re$, thus demonstrating that the PT narrative does not explain the origin of intermittency everywhere. More recent experiments also use a large resistance \cite{barkley2015rise}, and experiments with a constant mass flux \cite{darbyshire1995transition} have demonstrated convincingly that intermittency can also exist apart from PT's mechanism, although the typical method of instantaneously perturbing the flow renders the experimental initial conditions themselves intermittent. Rotta's insight laid the foundation for studying the patchy, localized turbulence, now believed to originate from special exact solutions of the governing Navier-Stokes equations such as nonlinear traveling waves \cite{mullin2011experimental,eckhardt2007turbulence}. Most recent work has focused as Rotta did on the vicinity of the critical $Re$ where non-expanding patches called ``puffs" dominate, or considered only instantaneous perturbations at higher $Re$ \cite{barkley2015rise}. Thus, with a few exceptions \cite{tritton2012physical,stassinopoulos1994periodic,fowler2003intermittency}, the PT mechanism has largely disappeared from any discussion of the transition \cite{barkley2016theoretical,mullin2011experimental,eckhardt2007turbulence}. However, this leaves neglected an important regime of transitional flow, a flow that transitions at $Re \gtrsim 3000$, and for which the pressure gradient and perturbation are constant.

%However, these studies have not shown that Prandtl and Tietjen's mechanism is irrelevant in all circumstances. 
%In most practical situations such as in industry, neither a large resistance nor a constant mass flux are or can be used. Likewise, the flow is not disturbed in the special way of laboratory experiments to produce intermittency artificially, but more likely to be triggered by a constant perturbation such as roughness \cite{cotrell2008instability}.
In this Letter we revisit PT's argument and look at the intermittency of transitional pipe flow under essentially constant conditions. It is driven by a constant pressure drop, disturbed continuously, and when not in the transition regime, the variation in $Re$, $\delta Re$, is small ($\delta Re / Re < 0.01$). We demonstrate the validity of the PT mechanism by developing a simple model based on their arguments that quantitatively reproduces the essential features of the intermittency in our experiments. Key to the success of our model is accounting for the external resistance, which we systematically vary, as well as accurately incorporating the growth of turbulent patches. The experiments and model together suggest a startling conclusion: under constant conditions and for $Re \gtrsim 3000$, there is always a regime of intermittency consistent with the PT argument.

% \begin{figure*}
% \centering
% \includegraphics[width=\textwidth]{introRoughExplanation2.eps}
% \caption{Plot of instantaneous friction $f$ vs. instantaneous non-dimensional flow speed $Re$. The friction is determined from the pressure drop over a finite section of the pipe near its end. The flow speed is determined using a flowmeter and is the same everywhere in the pipe since the flow is incompressible. Top inset: the friction $f$ as a function of time. Bottom inset: the Reynolds number $Re$ as a function of time. \textcolor{red}{This needs to be changed a lot. Put the average laminar and turbulent $f$ lines in c?}}
% \label{introRoughExplanation}
% \end{figure*}

\begin{figure}[t]
\centering
\includegraphics[width=1.0\linewidth]{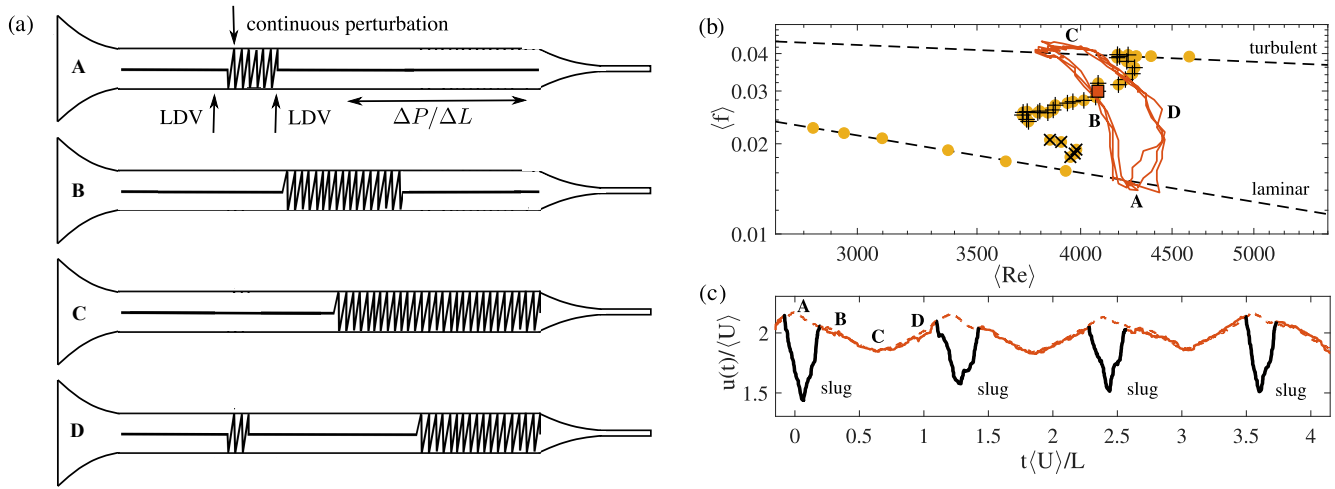}
\vspace{-2em}
\caption{(a) Schematic of the pipe experiment modelled after that appearing in \citet{reynolds1883an}, with a contracting entrance section and a final narrow pipe section for added resistance. Flow is from left to right. Straight horizontal lines indicate laminar flow, and jagged flow indicates a slug. The pressure measurement section for determining $\Delta P / \Delta L$ and thus $f$ in (b) is shown, as well as the two positions for measuring the velocity with LDVs in (c). The flow is disturbed continuously with an obstacle to set a transition $Re_C$. The flow state at points {\bf A}$\rightarrow${\bf D} from (b) and (c) are represented schematically. (b) Plot of average friction $\langle f \rangle$ vs. average $\langle Re \rangle$. The lower and upper curves are the laminar and turbulent friction curves, respectively. The transitional data are either periodic ($+$) or stochastic ($\times$). Overlaid on the mean data ($\circ$) we plot the instantaneous $f(t)-Re(t)$ curve for one periodic transitional data point ($\square$). This curve cycles clockwise through the points {\bf A}, {\bf B}, {\bf C}, and {\bf D}. (c) The normalized streamwise velocity $u(t)/\langle U \rangle$ at the centerline vs. the normalized time $t \langle U \rangle /L$ for $\square$, with {\bf A}$\rightarrow${\bf D} also denoted. The velocity was measured $100D$ downstream ($-$) and $10D$ upstream ($--$) from the perturbation. The shape of both curves is the same as $Re(t)$ but for the slugs (solid black line superimposed over $-$).}
\vspace{-1em}
\label{introRoughExplanation}
\end{figure}

For our experiments we carry out measurements of the flow rate, velocity, and the friction in a 2020 cm long, smooth, cylindrical glass pipe of diameter $D = 1$ cm $\pm$ 10 $\mu$m. The working fluid is water. Driven by gravity, the flow remains laminar up to $Re \approx 10000$. We restrict our attention to $3000 \lesssim Re \lesssim 7000$, for which the turbulent patches, called ``slugs" \cite{mullin2011experimental}, grow, an essential ingredient in the PT mechanism. We perturb the flow $\simeq 404D$ downstream (see Fig. \ref{introRoughExplanation}a) either continuously with an obstacle (a small $\simeq$ 0.63 mm diameter rod oriented perpendicular to the flow) or instantaneously with a syringe pump which injects a small amount of fluid from a 1 mm hole in the pipe wall. We denote by $L$ the distance from the perturbation to the end of the pipe. We can set a natural transition $Re$ when the flow becomes unstable, $Re_C$, by adjusting the rod protrusion. (This $Re_C$ should not be confused with the lower and universal critical $Re_C$ investigated by experiments of puff lifetimes \cite{avila2011onset}.)  We determine the instantaneous flow rate using a magnetic flowmeter (Yokogawa) and the total pressure drop $\Delta P_{\rm{tot}}$ by measuring the difference between the height of water surface in the source reservoir from the height of the water at the exit of the pipe, $\Delta h$. We also measure the instantaneous pressure drop in a $505D$ section that is $101D$ from the end of the pipe (see Fig. \ref{introRoughExplanation}a). Two Laser Doppler velocimeters (LDV, MSE) were also used to probe the flow (see Fig. \ref{introRoughExplanation}a). More experimental details can be found in the Supplementary Material ($SM$, Sec. II) and in Ref. \cite{cerbus2018laws}.

% figure aspect ratio (width/height) = 538/483 = 1.1139
% thus number of words is 150/1.1139+20 = 155

We begin by revisiting PT's mechanism through an examination of our experimental data for $Re$ and the non-dimensional friction factor $f = D \Delta P / \Delta L / (\rho U^2/2)$, where $\rho$ is the density, and $\Delta P$ is the pressure drop over a length $\Delta L$. We refer to Fig. \ref{introRoughExplanation}b, a traditional plot of $\langle f \rangle$ vs. $\langle Re \rangle$, to investigate the state of the system, where $\langle \rangle$ refers to the time-averaged value. As $\Delta P_{\textrm{tot}}$ slowly increases (via $\Delta h$), the data ($\circ$) initially conform to the lower laminar curve, but the flow becomes unstable due to the finite disturbance for $Re > Re_C \approx 4000$ (set by the obstacle) and the position of $\langle f \rangle - \langle Re \rangle$ deviates from the laminar curve thereafter. The first slugs appear stochastically ($\times$) \cite{zhang1994stochastic}, but this behavior spans only a narrow range of $\Delta P_{\rm{tot}}$. Thereafter the flow displays periodic behavior ($+$), which was the original focus of PT and thus ours as well.

In Fig. \ref{introRoughExplanation}b we plot the instantaneous $f(t)-Re(t)$ curve corresponding to one periodic data point ($\square$). To understand this curve, consider the point {\bf A} where the flow is laminar. Because $Re(t) > Re_C \simeq 4000$, a slug is created by the perturbation and begins to invade the flow, as indicated by a thick black line in Fig. \ref{introRoughExplanation}c (see also Fig. \ref{introRoughExplanation}a), and it expands aggressively as it is convected downstream \cite{mullin2011experimental}. The increased friction with $\Delta P_{\rm{tot}}$ = const. requires $Re(t)$ to decrease. The slug eventually reaches the pressure measurement section and partially fills it, raising the value of $f(t)$ to point {\bf B}, until the flow there is fully turbulent at point {\bf C} on the upper curve. As the turbulent patch leaves the pipe, $Re(t)$ increases and the flow's intermittency decreases, taking us through point {\bf D} (Fig. \ref{introRoughExplanation}a), until finally the flow is fully laminar again and we return to point {\bf A} to begin the cycle again. We now attempt to gain further insight by constructing a model to reproduce quantitative features.

We identify four essential ingredients, which we update and refine as necessary. The flow is driven by a constant pressure drop, $\Delta P_{\rm{tot}}$ ($\mathcal{I}1$), the pressure drop in a turbulent region is higher than a laminar one of the same length ($\mathcal{I}2$), slugs are convected and grow ($\mathcal{I}3$), and finally, a critical $Re_C$ is set by disturbing the flow continuously ($\mathcal{I}4$). We first combine $\mathcal{I}1$ and $\mathcal{I}2$ by distributing the constant $\Delta P_{\rm{tot}}$ between the laminar $\Delta P_{\rm{lam}}$ and slug $\Delta P_{\rm{turb}}$ portions of the flow. In addition, we also include the pressure drop of the system external to the experimental section, $\Delta P_{\rm{ext}}$, contributed by, for example, the entrance section. This gives the pressure drop balance $\Delta P_{\rm{tot}} = \Delta P_{\rm{lam}} + \Delta P_{\rm{turb}} + \Delta P_{\rm{ext}}$. As Fig. \ref{introRoughExplanation}b already indicates, when the pressure measurement region is laminar, $f(t)$ obeys the Hagen-Poiseuille law: $f_{\textrm{lam}} = 64/Re$, whereas when this region is turbulent, even during transition \cite{cerbus2018laws}, it obeys the empirical Blasius law: $f_{\textrm{turb}} = 0.3164Re^{-1/4}$. (This allows us to probe intermittency in a straightforward manner: the flow is intermittent if $f_{\textrm{lam}} < f(t) < f_{\textrm{turb}}$). We determine $\Delta P_{\rm{ext}}$ empirically in a series of experiments when the pipe is fully laminar, $\Delta P_{\rm{ext}} = \Delta P_{\rm{tot}} - \Delta P_{\rm{lam}}$ (see $SM$, Sec. I). Introducing the parameter $l$, the length of the pipe that is turbulent, results in (see $SM$, Sec. I):
\begin{equation}
    \frac{D^3 \Delta P_{\textrm{tot}}}{32 \rho \nu^2 L} = {\Big (} 1-\frac{l}{L} {\Big )} Re + B \frac{l}{L} Re^{7/4} + R(Re),
    \label{eq:mainEq}
\end{equation}
where $B = 0.3164/64$ is a constant combining the constants from the Hagen-Poiseuille and Blasius laws and $R = \frac{D^3 \Delta P_{\textrm{ext}}}{32 \rho \nu^2 L}$ is the normalized external resistance. The terms on the right hand side are the pressure drop contributions from the laminar ($\propto 1-l/L$), turbulent ($\propto l/L$) and external portions of the pipe, respectively. Previous work that split $\Delta P_{\textrm{tot}}$ between a laminar and turbulent contribution also predicted oscillations, but they were unable to show quantitative agreement between model and experiment \cite{stassinopoulos1994periodic,fowler2003intermittency}. This highlights the importance of accounting for the external resistance $R$ and accurately incorporating slug growth rates, both of which were not included in these approaches.

\begin{figure}
\centering
\includegraphics[width=\linewidth]{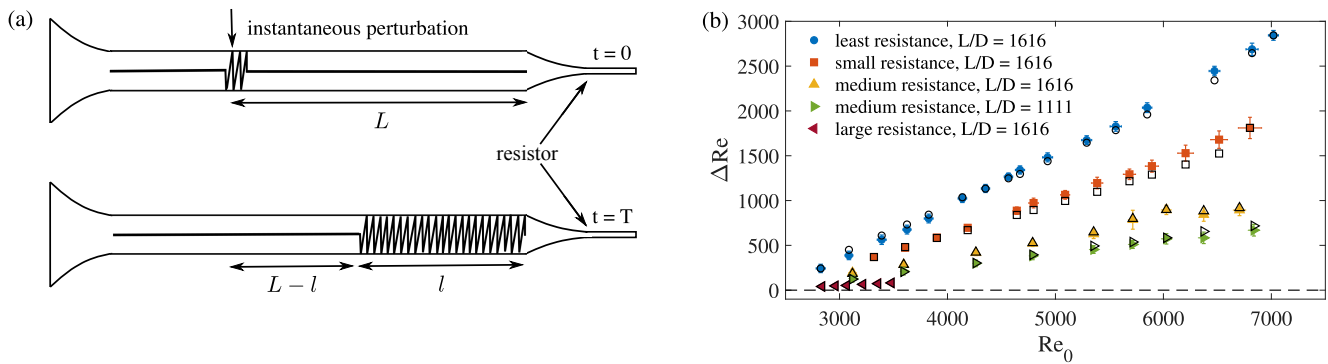}
\vspace{-2em}
\caption{(a) Schematic of transitional pipe flow as in Fig. \ref{introRoughExplanation}. A single slug is engendered at $t=0$ by a perturbation, which grows to its maximum size when it reaches the end of the pipe at $t = T$. (b) Plot of deviation in $Re$, $\Delta Re$, vs. the initial $Re_0$ for several resistances $R$. Increasing $R$ or decreasing $L$ reduces $\Delta Re$. A dependence on $L$ is to be expected and is also seen in work on pulsatile flow \cite{xu2018effect}. The predicted $\Delta Re$, open symbols, are in excellent accord with the experimental data, filled symbols.}
\vspace{-1em}
\label{Re0DeltaRe}
\end{figure}
% figure aspect ratio (width/height) = 552/477 = 1.1572
% thus number of words is 150/1.1572+20 = 150

As a first step in validating our refined model, we use Eq. \ref{eq:mainEq} to predict the maximum change in $Re$ when a single slug is created, utilising $\mathcal{I}1-\mathcal{I}3$. We perform experiments in which we systematically vary $R$ by adding short sections of smaller diameter pipes (see Fig. \ref{Re0DeltaRe}a), a ``resistor", to the pipe system \cite{samanta2011experimental,de2009experimental} and determine the $Re$ dependence of $R$ empirically (see $SM$, Sec. I). We then perturb the flow instantaneously at a distance $L$ from the end of the pipe where the laminar flow is fully developed. We adjust $\Delta P_{\textrm{tot}}$ via $\Delta h$ to set an initial $Re = Re_0$ and seek the maximum deviation from $Re_0$: $\Delta Re = Re_0 - Re_{\textrm{min}}$, where $Re_{\textrm{min}}$ is the minimum $Re$. For each $\Delta h$ and $R(Re)$ we perform the experiment at least three times to determine averages and uncertainties. For constant $\Delta P_{\rm{tot}}$, we can write Eq. \ref{eq:mainEq} at both $Re_0$ and $Re_{\rm{min}}$ and equate them to show that:
% \begin{multline}
\begin{equation}
    Re_0 + R(Re_0) = \frac{D^3 \Delta P_{\textrm{tot}}}{32 \rho \nu^2 L} =    {\Big (} 1-\frac{l}{L} {\Big )} Re_{\textrm{min}} + B \frac{l}{L} Re_{\textrm{min}}^{7/4} + R(Re_{\textrm{min}}),
    \label{eq:meanEquation}
\end{equation}
% \end{multline}
where for $Re = Re_0$, $l = 0$ by definition. The $l/L$, which we next estimate, also depends on $Re$. We suppose that the minimum value $Re_{\textrm{min}}$ occurs when $l/L$ is at its maximum as the growing slug reaches the end of the pipe. The maximum $l/L$ can be estimated using the slug front speed, $u_F$, and back speed, $u_B$. If $T$ is the time it takes the slug front to reach the end of the pipe, then $L = u_F T$ and $L-l = u_B T$, which can be rearranged to find $l/L = (u_F - u_B)/u_F$. We made our own estimates of $u_F$ and $u_B$ (see $SM$, Sec. II) because the literature values are for practically constant $Re$ \cite{lindgren1969propagation,wygnanski1973transition,nishi2008laminar,barkley2015rise}. Because the external resistance in these experiments is deliberately smaller, the $Re$ here is not constant. We then solve Eq. \ref{eq:meanEquation} numerically, and Fig. \ref{Re0DeltaRe}b shows that its predictions are in excellent accord with the experimental results. The variation in the $Re$ as the slug grows also leads to a subtle dependence on the pipe length $L$, as the growing slug has more time to slow down the flow if $L$ is larger. Thus as Fig. \ref{Re0DeltaRe}b shows, for the same external resistance but smaller $L/D$, $\Delta Re$ is smaller.

%Moreover, since our largest resistance is serendipitously comparable to our estimate of Rotta's external resistance (see $SM$), our results suggest that if he probed $Re \gtrsim 3000$, he would have observed fluctuations in $Re$.

\begin{figure}
\centering
\includegraphics[width=\linewidth]{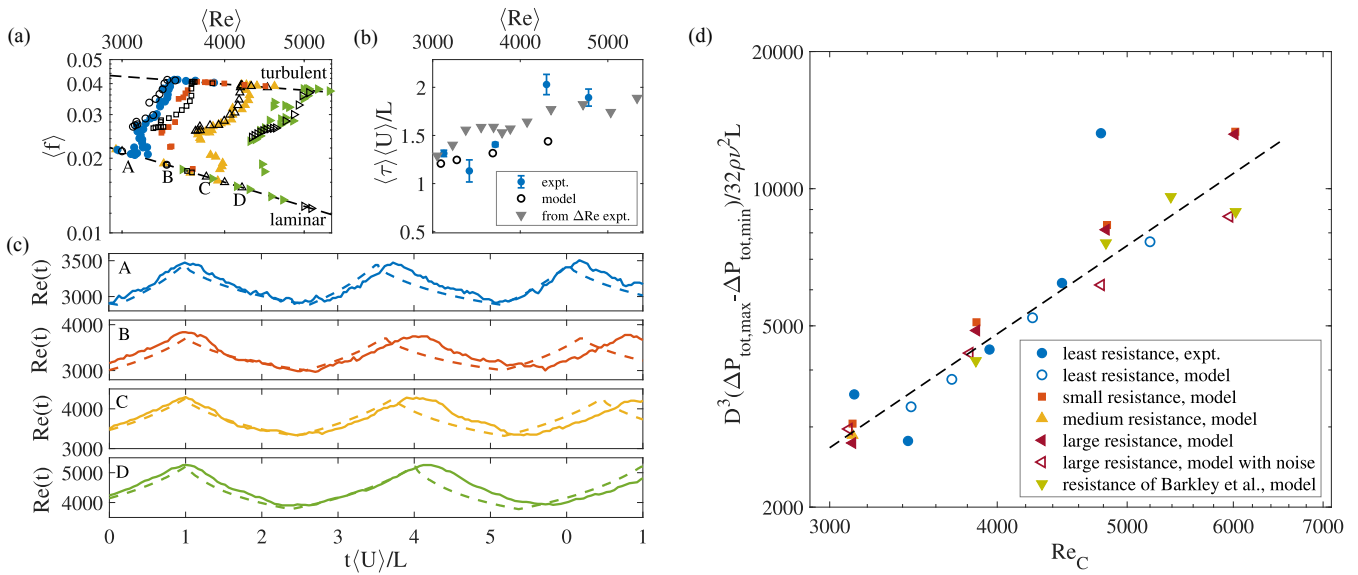}
\vspace{-2em}
\caption{(a) experiments, filled symbols, and model predictions, open symbols, for $\langle f \rangle$ vs. $\langle Re \rangle$. The model data closely follows the experimental data. For each series of data, $A$-$D$, we plot in (c) below the normalized time series of $Re(t)$ vs. $t\langle U \rangle /L$, choosing the point of minimum $\langle Re \rangle$ in the periodic intermittent regime. The amplitudes and normalized periods $\tau\langle U \rangle/L$ closely match the experiments. (b) $\tau\langle U \rangle/L$ determined from experiments, the model, and from the previous $\Delta Re$ experiments (Fig. \ref{Re0DeltaRe}, $\tau\langle U \rangle/L \simeq \langle U \rangle/u_B$). The normalized periods differ by $\lesssim 5 \%$ in most cases. (d) Non-dimensional intermittency span vs. transition $Re_C$ for experiments and model simulations ($R$ as in Fig. \ref{Re0DeltaRe}). We also include model simulations using an extremely large $R$ estimated from Barkley et al.'s experiments \cite{barkley2015rise} (see $SM$, Sec. V for details). The ratio of smallest to largest resistance is $\simeq 500$ and the $L/D$ varied up to $\simeq 24 \%$. We also probed the effect of external noise by adding normally distributed noise with zero mean and standard deviation of 0.1 to $Re(t)$ at each step in the integration of Eq. \ref{eq:timeDepEq}. Despite these differences, all data collapse onto a common curve and exhibit a non-negligible span of intermittency that increases with $Re_C$ ($\propto Re_C^2$, $--$). }
\vspace{-1.5em}
\label{oscillationsFrictionInsetIntermittencySpan}
\end{figure}
% figure aspect ratio (width/height) = 541/490 = 1.1041
% thus number of words is 150/1.1041+20 = 156

We now proceed to develop a time-dependent version of the model to reproduce the oscillations, now incorporating a critical $Re_C$ ($\mathcal{I}4$). We take the time derivative of Eq. \ref{eq:mainEq} ($\mathcal{I}1$, $\mathcal{I}2$), subject to the constraint $\Delta P_{\textrm{tot}} =$ const. ($\mathcal{I}1$), which yields:
\begin{equation}
    \frac{d Re}{dt} = \frac{\frac{d(l/L)}{dt} {\Big (} Re - B Re^{7/4} {\Big )}}{{\Big (} 1-\frac{l}{L} {\Big )} + \frac{7}{4} B Re^{3/4} {\Big (} \frac{l}{L} {\Big )} + \frac{d R}{d Re}}.
    \label{eq:timeDepEq}
\end{equation}
To determine the time-dependence of $l/L$ we use a recent model which has had significant success in reproducing the growth rates ($\mathcal{I}3$) of slugs \cite{barkley2015rise}. The complexity of slug growth is reduced to two coupled partial differential equations for a variable representing the turbulence intensity, $q$, and the pipe centerline velocity $u$. Now together with Eq. \ref{eq:timeDepEq} we have a set of coupled partial and ordinary differential equations. Since the $l/L$ in Eq. \ref{eq:timeDepEq} is simply the total turbulent fraction, we do not use the spatial information of the partial differential equations in Eq. \ref{eq:timeDepEq}. This system of equations is similar to, but simpler than, the systems of coupled differential equations used to model arterial flow \cite{reymond2009validation}.

We perform several experiments without an external resistor, although $R \neq 0$, systematically changing the transition $Re_C$ by adjusting the amplitude of the perturbation ($\mathcal{I}2$). For each $Re_C$, set by adjusting the obstacle, we repeated the experiment of Fig. \ref{introRoughExplanation}b, slowly increasing $\Delta P_{\textrm{tot}}$ to take the system from laminar, to intermittent, to turbulent (see Fig. \ref{oscillationsFrictionInsetIntermittencySpan}a). For our model, we integrate Eq. \ref{eq:timeDepEq} along with the coupled partial differential equations from Barkley et al.'s model \cite{barkley2015rise}, which we transformed into laboratory units (see $SM$, Sec. III). To reproduce the behavior in our experiments we add a constant perturbation to the Barkley model, the amplitude of which we varied to set different transition $Re_C$ as in the experiments ($\mathcal{I}4$). This deterministic model is not able to reproduce the initial region of intermittency, in which slugs appear stochastically, but it both quantitatively reproduces the oscillations and the shapes of the $\langle f \rangle$ vs. $\langle Re \rangle$ curves (see Fig. \ref{oscillationsFrictionInsetIntermittencySpan}c).

% \begin{figure}
% \centering
% \vspace{-1em}
% \includegraphics[width=\linewidth]{intermittencySpanUpdated4.eps}
% \vspace{-2em}
% \caption{Non-dimensional intermittency span vs. critical $Re_C$ for experiments and model simulations ($R$ as in Fig. \ref{Re0DeltaRe}). We also include model simulations using an extremely large $R$ estimated from Barkley et al.'s experiments \cite{barkley2015rise} (see $SM$, Sec. V for details). The ratio of smallest to largest resistance is $\simeq 500$. We also probed the effect of external noise by adding normally distributed noise with zero mean and standard deviation of 0.1 to $Re(t)$ at each step in the integration of Eq. \ref{eq:timeDepEq}. Despite these differences, all data collapse onto a common curve and exhibit a non-negligible span of intermittency that increases with $Re_C$ ($\propto Re_C^2$, $--$).}
% \vspace{-1.5em}
% \label{intermittencySpan}
% \end{figure}
% % figure aspect ratio (width/height) = 555/460 = 1.2065
% % thus number of words is 150/1.2065+20 = 144

As Figs. \ref{Re0DeltaRe}, \ref{oscillationsFrictionInsetIntermittencySpan}a, \ref{oscillationsFrictionInsetIntermittencySpan}b, \ref{oscillationsFrictionInsetIntermittencySpan}c show, our model, based on the PT mechanism, is in excellent accord with the experimental data. We now use this result to demonstrate the generality of the PT mechanism. As already noted, Rotta tested the PT mechanism by restricting attention near the universal critical point ($Re \lesssim 3000$) and by increasing the external resistance. The former invalidates the PT mechanism because it removes slug growth, an essential ingredient ($\mathcal{I}3$). As Fig. \ref{Re0DeltaRe}b demonstrates, the latter approach of increasing $R$ unsurprisingly reduces deviations in $Re$ ($\Delta Re$). Indeed, this principle is broadly used to maintain a nearly constant $Re$ in constant pressure gradient transitional pipe flow experiments. If fluctuations can be completely eliminated, one would expect no intermittency and thus in our $\langle f \rangle - \langle Re \rangle$ curves there would be a discontinuous jump from the laminar to turbulent friction curves at $Re_C \gtrsim 3000$. To test the hypothesis that the regime of intermittency shrinks as $R$ increases, we plot versus $Re_C$ in Fig. \ref{oscillationsFrictionInsetIntermittencySpan}d the normalized difference between the pressure drop at the end of the intermittent regime $\Delta P_{\rm{tot,max}}$ and at the beginning $\Delta P_{\rm{tot,min}}$. Despite spanning over two orders of magnitude in $R$, and even in the presence of noise, all data collapse onto a single curve that inexplicably increases with $Re_C$. When expressed in terms of the true control variable, the normalized pressure gradient, the intermittency span is independent of $R$. Moreover the attendant intermittency is not negligible, since the fraction of flow filled by patches necessarily advances continuously from zero to unity as the pressure drop is increased from $\Delta P_{\rm{tot,min}}$ to $\Delta P_{\rm{tot,max}}$. However, we note that while the intermittency is substantial, the relative magnitude of the fluctuations in $Re$ can be substantially reduced by increasing $R$, as shown in Fig. \ref{Re0DeltaRe}b. Near the natural transition point $Re_C$ the finite-amplitude threshold is very sharp and thus very sensitive for $Re \gtrsim 3000$ \cite{tasaka2010folded,hof2003scaling}, so that even these small variations in $Re$ are sufficient for the PT mechanism to function. Prandtl-Tietjens intermittency is thus an intrinsic feature of continuously perturbed and constant pressure driven flows, for which substantial intermittency and tunable fluctuations in $Re$ are unavoidable if $Re > Re_C > 3000$.

In conclusion, we have developed a model inspired by Prandtl and Tietjens' classic argument that is in excellent quantitative agreement with experiments. Essential to the model's success was accurately accounting for the external resistance and slug growth rates. We began our inquiry by noting that, beginning with Rotta \cite{rotta1956experimenteller}, the Prandtl-Tietjens argument has been considered irrelevant. Together, our experiments and model suggest that intermittency engendered by the Prandtl-Tietjen mechanism is in fact an intrinsic feature of constant pressure driven pipe flow for constant conditions, and for $Re \gtrsim 3000$. Rotta did not avoid it by increasing the resistance in his pipe, which ultimately cannot remove the intermittency engendered by the PT mechanism (Fig. \ref{oscillationsFrictionInsetIntermittencySpan}d), but by restricting attention to $Re \lesssim 3000$ \cite{rotta1956experimenteller}, just as many other laboratory experiments restrict attention to $Re < Re_C$ \cite{avila2011onset,barkley2015rise,nishi2008laminar} in order to consider the effect of instantaneous perturbations. Thus while the PT mechanism elucidated here does not apply to those important studies, neither do they directly address the intermittency in the early experiments of Hagen \cite{hagen1839ueber}, Brillouin \cite{brillouin1907leccons}, and others \cite{letellier2017intermittency}, or those conducted here. Most pipes will have a natural transition $Re_C$ set by imperfections such as wall roughness \cite{cotrell2008instability,tao2009critical}, and here it is the Prandtl-Tietjens mechanism which provides the route to turbulence. Fusing old insights \cite{tietjens1957applied} and new \cite{barkley2015rise,cerbus2018laws} has broadened the impact of both, yielding new and practical understanding of transitional pipe flow.

\emph{Acknowledgments.} We thank Tom Mullin for suggesting this problem, as well as Pinaki Chakraborty and Hamid Kellay for helpful discussions. We thank an anonymous referee for correcting our interpretation of Rotta’s paper in an earlier draft. We thank the Service Informatique at the Laboratoire Ondes et Mati\`{e}re d'Aquitaine for computational support. R.T.C. gratefully acknowledges the support of a Marie Sk\l odowska-Curie Action Individual Fellowship (MSCAIF), and the support of the Okinawa Institute of Science and Technology (OIST) where the experiments were carried out.

% For conclusion:
% \begin{itemize}
% \item Negative statement: Without the Prandtl and Tietjen argument, intermittency at these $Re$ is only possible if the perturbation is itself intermittent, for example if the flow is perturbed by a jet for finite time.
% \item Positive statement: There is always a region of $Re$ for which Prandtl and Tietjen's argument is valid.
% \end{itemize}

% \vspace{2em}
% \noindent {\bf Thoughts: }

% There seems to be a period doubling that can occur? The minimum Re in the oscillation set has the largest amplitude and one slug at a time.

% Some additional things: \\
% (3) I can fit the $\Delta Re$ vs. $Re$ fairly well (see notebook pg. 52-53)
% (4) I need to show that for fixed $L$, $Re_C$ ($=Re_0$ for the single perturbation experiments), the amplitude $\Delta Re$ decreases with the resistance, but not enough!
% (5) Mullin was right: the amplitude follows a square root law from turbulence to the oscillation, whereas it is intermittent from laminar (and discontinuous for the simulations, which have no noise)

% \bibliographystyle{apsrev}
\bibliography{pipeIntermittency}

\end{document}